# Blood-derived lncRNAs as potential biomarkers for early cancer diagnosis:

# The Good, the Bad and the Beauty

Cedric Badowski[1#], Bing He[2], Lana X Garmire[1,2, #]

[1] Previous address: University of Hawaii Cancer Center, Epidemiology, 701 Ilalo Street, Honolulu, HI 96813, USA

[2] Current address: Department of Computational Medicine and Bioinformatics, University of Michigan, MI, 48105, USA

[#] Corresponding authors:

Cedric Badowski: cedricbadowski@gmail.com

Lana X. Garmire: LGarmire@med.umich.edu

**Abstract**

Cancer ranks as one of the deadliest diseases worldwide. The high mortality rate associated with cancer is partially due to the lack of reliable early detection methods and/or inaccurate diagnostic tools such as certain protein biomarkers. Cell-free nucleic acids (cfNA) such as circulating long non-coding RNAs (lncRNAs) have recently been proposed as a new class of potential biomarkers that could improve cancer diagnosis. The reported correlation between circulating lncRNA levels and the presence of tumors has triggered a great amount of interest among clinicians and scientists who have been actively investigating their potentials as reliable cancer biomarkers. In this report, we review the progress achieved ("the Good") and challenges encountered ("the Bad") in the development of circulating lncRNAs as potential biomarkers for early cancer diagnosis. We report and discuss the specificity and sensitivity issues of blood-based lncRNAs currently considered as promising biomarkers for various cancers such as hepatocellular carcinoma, colorectal cancer, gastric cancer and prostate cancer. We also emphasize the potential clinical applications ("the Beauty") of circulating lncRNAs both as therapeutic targets and agents, on top of diagnostic and prognostic capabilities. Based on different published works, we finally provide recommendations for investigators who seek to investigate and compare the levels of circulating lncRNAs in the blood of cancer patients compared to healthy subjects by RT-qPCR or Next Generation Sequencing.

**Introduction**

Cancer ranks as one of the deadliest diseases worldwide. Despite ongoing efforts to develop new treatments and a better understanding of the mechanisms underlying tumorigenesis, it remains difficult to treat cancers, particularly when diagnosed at late stages with a poor prognosis. The high mortality rate associated with cancer is partially due to the lack of early detection methods and/or inaccurate diagnostic tools such as certain protein biomarkers.

Protein or peptide-based biomolecules such as glycoproteins constitute most of the currently available cancer biomarkers. Variations in their levels in tissues or blood may indicate the development of diseases such as cancer. Protein markers can be detected in tissue biopsy sections analyzed by immunohistochemistry (IHC) upon diagnostic notably to determine cancer molecular subtype. For instance, breast tumor tissues are commonly assessed for the presence of estrogen receptor (ER) to determine their ER-positive or ER-negative status. However, some protein biomarkers are reportedly unreliable as they generate a significant amount of false-positive and/or false-negative results. Plasma alpha fetoprotein (AFP), one of the most frequently used biomarkers for diagnosis of hepatocellular carcinoma [1], has been described by many as a marker with low sensitivity and/or specificity [2-5]. Conventional serological biomarkers such as carbohydrate antigen 153 (CA153), cancer antigen 125 (CA125), CA27.29 and carcinoembryonic antigen (CEA) remain controversial due to poor specificity and sensitivity [6-11].

The poor reliability of certain protein biomarkers is partially due to the nature of the biomarker itself. The detection of proteins and peptides indeed relies on the use of antibodies that may or may not be specific to the desired marker as the epitope recognized by the antibodies may be present on other tissue components [12]. Unreliable antibodies currently represent a major issue in biomedical research in general and can significantly comprise the outcome of a study or diagnosis. Another issue with traditional histology analyses is the need for actual tissue biopsies. This invasive and inconvenient technique may discourage potential cancer patients to proceed with the entire diagnosis procedure. Thus, the development of non-invasive non-protein biomarkers is currently needed.

Cell-free nucleic acids (cfNA) or circulating nucleic acids (CNAs) have recently been proposed as a new class of potential biomarkers that could improve cancer diagnosis [13]. CNA PCA3 (prostate cancer associated 3) has notably been approved by the FDA and is currently being sold as Progensa by Hologic Gen Probe (Marlborough, MA, USA) for the diagnosis of prostate cancer [14-16]. Circulating long non-coding RNAs or lncRNAs (non-coding RNAs of 200 nucleotides or more) such as PCA3 seem more reliable than other CNAs due to their high stability in the bloodstream and poor sensitivity to nuclease-mediated degradation. Arita et al. especially showed that plasmatic lncRNAs are resistant to degradation induced by repetitive freeze-thaw cycles, as well as prolonged exposure to 45C and room temperatures [17]. The stability of lncRNAs in the bloodstream appears to originate from the presence of extensive secondary structures [18], the transport by protective exosomes [19] as well as stabilizing posttranslational modifications.

Circulating lncRNAs are, directly or indirectly, correlated with the presence of tumors in vivo. Tumorigenesis has indeed been reported to be associated with changes in levels of circulating lncRNAs. For instance, the blood of patients with hepatocellular carcinoma was shown to contain elevated levels of lncRNA HULC (for "highly up-regulated in liver cancer") [20]. Moreover, HULC, H19, HOTAIR and GACAT2 (for "gastric cancer associated transcript 2") were found to be significantly increased in the plasma of gastric cancer (GC) patients compared to healthy individuals [21-24]. Higher levels of circulating lncRNAs GIHCG (for "gradually increased during hepatocarcinogenesis") and ARSR (for "activated in RCC with sunitinib resistance") were reported in renal cell carcinoma (RCC) patients [25-27]. LncRNA MALAT-1 (metastasis associated lung adenocarcinoma transcript 1) was detected in significantly higher quantities in the plasma of patients with prostate cancer as compared to healthy subjects [28]. This study showed that tumors were at the origin of MALAT-1 variations, since the surgical removal of the cancerous tissues induced a dramatic reduction in circulating MALAT-1, while plasmatic levels of this lncRNA increased upon ectopic implantation of a tumoral xenograft in mice [28]. Similarly, levels of circulating lncRNAs GIHCG and ARSR significantly dropped after resection of RCC tumors, while plasma levels of H19, A174084 and GACAT2 markedly decreased in GC patients postoperatively, further supporting a direct correlation between abnormal levels of circulating lncRNAs and tumorigenesis [21, 24-26, 29, 30]. In the last few years, increasing numbers of reports have emphasized the correlation between various types of cancers and changes in systemic levels of specific lncRNAs. Consequently, clinicians and scientists have been actively investigating their potential as reliable cancer biomarkers. Some of these circulating lncRNAs have shown greater diagnostic performance than conventional glycoprotein markers. Plasmatic H19, for instance, was reported as more reliable than carcinoembryonic antigen (CEA) and carbohydrate antigen 153 (CA153) for the diagnosis of breast cancer [31]. Likewise, a serum three-lncRNA signature consisting of PTENP1, LSINCT-5 and CUDR (also known as UCA1) significantly outperformed CEA and CA19-9 in gastric cancer diagnostic studies [32].

In this report, we review the progress achieved and challenges encountered in the development of circulating lncRNAs as potential biomarkers for early cancer diagnosis. We report and discuss the specificity and sensitivity of blood-based lncRNAs currently considered as promising biomarkers for various cancers such as hepatocellular carcinoma, colorectal cancer, gastric cancer and prostate cancer. We also highlight potential therapeutic applications for circulating lncRNAs both as therapeutic targets and agents, on top of diagnostic and prognostic purposes. Based on recommendations from different published works, we finally provide recommendations for investigators who seek to investigate and compare the levels of circulating lncRNAs in the blood of cancer patients compared to healthy subjects by RT-qPCR or Next Generation Sequencing.

**Blood-based lncRNAs as potential circulating biomarkers for cancer diagnosis**

*Changes in circulating lncRNA levels specifically correlate with cancer development*

Most studies focusing on circulating lincRNAs have been initiated based on prior observations reporting changes in lncRNA levels in cancer tissue samples. For instance, MALAT-1 was first shown to be upregulated in various cancer tissues including lung and prostate tumors [33, 34]. Using peripheral blood cells as a lincRNA source for their study, Weber et al. later showed that MALAT-1 levels could reflect the presence of non-small-cell lung cancer with a specificity of 96% [35] (Table 1). Analysis of MALAT-1 levels in the plasma of prostate cancer patients and healthy subjects revealed that changes in circulating MALAT-1 levels correlated with prostate cancer with relatively high specificity (84.8%) [28]. LncRNA GIHCG was originally found to be upregulated in cancer tissue samples from HCC and RCC tumors [26, 36]. GIHCG was also detected at higher levels in the serum of RCC patients, suggesting not only the existence of a robust correlation between tissue-based and circulating levels of GIHCG but also the likelihood that the variations in serum levels of GIHCG originate from the tumoral tissue itself [26]. Moreover, serum GIHCG levels were able to distinguish RCC patients from healthy individuals with a specificity of 84.8%. Other lncRNAs have been reported to detect various cancer types with relatively high specificity. For instance, HOTAIR (HOX transcript antisense RNA) has shown high efficacy in identifying samples from colorectal cancer patients with a specificity of 92.5% [37]. Changes in plasmatic levels of lncRNA LINC00152 were found to correlate with gastric cancer with a specificity of 85.2% [19] (Table 1). LNC00152 has also been suggested as a reliable blood-based biomarker for hepatocellular carcinoma (HCC) [38]. The high prevalence of HCC in certain parts of the world such as Asia or Africa is undeniably alarming and it has become a major public health matter in many countries. Reliable biomarkers are desperately needed to detect this deadly cancer at an early stage. Many circulating lncRNAs have shown a significant correlation with HCC and represent promising candidates for HCC diagnostic applications (Table 1). Several studies from Egypt identified lncRNA-UCA1 as a potential serum-based biomarker for the detection of HCC. The specificities obtained were 82.1% [39] and 88.6% [40]. These studies also reported WRAP53 and CTBP as potential biomarkers for HCC with a specificity of 82.1% [39] and 88.5% [40], respectively. In Asia, Jing et al showed that lncRNA SPRY4-IT1 represents another promising blood-based biomarker for the diagnosis of hepatocellular carcinoma [41].

Many more circulating lncRNAs have been proposed as potential blood-based biomarkers for HCC and other types of cancers with relatively high specificity (Table 1).

*Poor specificity or sensitivity of lncRNA biomarkers. Challenges and potential impacts on diagnosis*

The diagnostic power of circulating biomarkers has yet to reach its maximum potential. Indeed, the diagnostic performance of many circulating lncRNAs remains relatively poor when taken individually. Several lncRNAs reportedly have either poor sensitivity or poor specificity towards a specific cancer type. For instance, plasmatic levels of HULC were able to detect gastric cancer samples with a sensitivity of only 58% [22] (Table 1). Similarly, MALAT-1 has shown a sensitivity of only 58,6% when testing plasma samples from prostate cancer patients and healthy subjects. This moderate sensitivity implies that the use of MALAT-1 as a blood-based prostate cancer biomarker may result in a significant number of false-negative results, as actual cancer samples may not be detected. MALAT-1 has also been investigated as a potential biomarker for non-small-cell lung cancer [35, 42]. However, with a sensitivity of only 56%, MALAT-1 may also face multiple challenges before becoming a reliable blood-based biomarker for lung cancer diagnosis (Table 1). One unsolved issue is notably the reported lack of correlation between the levels of circulating MALAT-1 in lung cancer patients and the levels of this lncRNA in lung cancer tissues. Indeed, the comparative analysis of whole blood samples from 105 lung cancer patients and 65 healthy subjects revealed a decrease in blood MALAT-1 levels in cancer patients, while lung cancer tissues showed higher MALAT1 expression [42]. The lack of strong sensitivity and the poor correlation between tissue and blood levels may arise from the fact that MALAT-1 is reportedly undergoing a certain degree of degradation in the bloodstream [28]. One of the resulting fragments has notably been referred to as MD-mini RNA (for metastasis associated in lung adenocarcinoma transcript 1 derived miniRNA) [28].

The degradation of MALAT-1 in the bloodstream may not be an isolated case and, probably, many more lncRNAs are actively being degraded once they enter the circulation. Degradation of circulating lncRNAs may increase in cancer patients as several studies reported that tumorigenesis is often associated with higher RNAse activity in the bloodstream [43]. In fact, long before circulating lncRNAs were considered as potential cancer biomarkers, increased RNAse activity in the serum of cancer patients was suggested as a mean of early cancer detection [44, 45]. In their study, Reddi and Holland notably reported that 90% of the patients with pancreatic cancer showed a dramatic increase in serum RNAse levels (above 250 units / mL).

They hence promoted the use of high serum RNAse activity as a biomarker for pancreatic carcinoma. Other cancers such as chronic myeloid leukemia have also been reported to be associated with a higher level of plasmatic RNAse activity [46]. RNAses circulating in the bloodstream notably constitute cytotoxic agents secreted by immune cells as part of anti-cancer defense mechanisms that aim at lysing transformed cells by activating cell death pathways [47]. For instance, an RNAse secreted by human eosinophils is known to induce the specific apoptosis of Kaposi's sarcoma cells without affecting normal human fibroblasts [48]. RNAse L was shown to suppress prostate tumorigenesis by initiating a cellular stress response that leads to cancer cell apoptosis [49, 50]. Tumors, on the other hand, reportedly display lower RNAse activity to promote protein synthesis and cell proliferation [43]. The reported difference in RNAse activity in tumors versus circulation may explain seemly paradoxical data when comparing lncRNA levels in tissues and blood such as in the case of MALAT1. While many studies have shown positive correlations between tissue and blood lncRNAs, the reported increased RNAse activity in the blood of some cancer patients may promote the degradation of circulating lncRNAs to a degree that would depend on the nature of cancer and/or lncRNA studied. This could represent a significant challenge for investigators as RT-qPCR analyses may not detect fragments of an investigated lncRNA possibly compromising the outcome of a study.

LINC00152 is another circulating lncRNA that has been actively investigated as a potential cancer biomarker. However, LINC00152 has shown a sensitivity of only 48.1% when analyzing plasma samples from gastric patients and healthy subjects, limiting its diagnostic performance as well (Table 1). It is currently not clear if LINC00152 is undergoing degradation in the bloodstream. Other circulating lncRNAs have shown poor specificity in the detection of specific cancers. For instance, GACAT2 reportedly has a specificity of only 28% when comparing plasma samples from gastric cancer patients and healthy subjects [21], while several studies have shown that H19 is capable of detecting samples from gastric cancer patients with a specificity of only 58 % [17] or 56.67% [51] (Table 1). This implies that diagnosis based on the quantification of plasmatic levels of H19 or GACAT2 may potentially result in a significant number of false-positive results when testing for gastric cancer. It is also the case for lncRNA SPRY4-IT1 regarding the diagnosis of hepatocellular carcinoma (HCC) with a specificity of only 50% (Table 1).

Therefore, significant improvements are required before most individual circulating lncRNAs become reliable blood-based cancer biomarkers.

*Combination of circulating lncRNAs for greater diagnostic performance*

To compensate for the moderate specificity/sensitivity of certain circulating lncRNAs and increase their diagnostic performance, several studies have combined the diagnostic values of several circulating lncRNAs. For instance, Hu et al., integrated lncRNAs SPRY4-IT1, ANRIL and NEAT1 in their studies on non-small-cell lung cancer and obtained a specificity of 92.3%, a sensitivity of 82.8%, and an AUC (ROC) (area under the ROC curve - receiver operating characteristic) of 0.876 [52] (Table 1). When combined with POU3F3 and HNF1AAS1, SPRY4-IT1 displayed a sensitivity of 72.8% and a specificity of 89,4% (AUC: 0.842) in the detection of esophageal squamous cell carcinoma [53]. Yu et al reported that the combination of circulating lncRNAs PVT1 and uc002mbe.2 reflected the presence of hepatocellular carcinoma with a specificity of 90.6% and a sensitivity of 60.5% [54]. The integrated analysis of plasmatic levels of XLOC_006844, LOC152578 and XLOC_000303 allowed the detection of colorectal cancer with a specificity of 84%, a sensitivity of 80% and an AUC of 0.975 [55]. Other examples include the combination of lncRNAs RP11-160H22.5, XLOC_014172 and LOC149086 which produced a sensitivity of 82% and a specificity of 73% (AUC: 0.896) for the diagnosis of hepatocellular carcinoma [3] (Table 1). Some studies have investigated the diagnostic signature of more than 3 circulating lncRNAs. Zhang et al for instance identified a panel of five plasma lncRNAs (BANCR, AOC4P, TINCR, CCAT2 and LINC00857) that was able to discriminate GC patients from healthy controls with an AUC of 0.91, outperforming CEA biomarker [56]. Wu et al. have reported that a 5-lncRNA signature could accurately distinguish serum samples of patients with renal cell carcinoma (RCC) from those of healthy subjects [57]. The combination of lncRNA-LET, PVT1, PANDAR, PTENP1 and linc00963 identified RCC samples with an AUC of 0.823. Each of these 5 lncRNAs was not individually capable of performing as well as the 5-lncRNA signature. PVT1 and PANDAR have also been investigated as part of a 8-lncRNA signature in plasma samples of patients with pancreatic ductal adenocarcinoma [58]. The 8-lncRNA signature was identified by using a custom nCounter Expression Assay (Nanostring Technologies, USA) that allows multiplex qPCR analyses using TaqMan probes.

Therefore, the signature generated by the combination of several blood-based lncRNAs reportedly provides better diagnostic performance than most individual circulating lncRNAs.

**Circulating lncRNAs as potential blood-based biomarkers for cancer prognosis**

Besides being potential blood-based biomarkers for early cancer diagnosis, circulating lncRNAs may also constitute valuable prognosis markers. Most studies assessing the ability of lncRNAs to predict disease evolution and eventual clinical outcome have been performed on cancer tissue samples [59-61]. However, a few studies based on the analysis of blood-derived samples indicate that circulating lncRNAs may also be able to reflect cancer prognosis. For instance, changes in plasmatic levels of lncRNAs XLOC_014172 and LOC149086 can distinguish metastatic HCC from non-metastatic HCC with a specificity of 90%, a sensitivity of 91% and an AUC of 0.934 (combined) [3]. HOTAIR can also be used as a negative prognostic marker for colorectal cancer with a sensitivity of 92,5%, a specificity of 67% and an AUC of 0.87 [37]. Moreover, lncRNA GIHCG has been proposed as a potential prognostic biomarker for renal cell carcinoma [26]. The 5-lncRNA signature reported by Wu et al., was also capable of discriminating benign renal tumors from metastatic renal cell carcinoma [57]. Similarly, the 8-lncRNA signature recently described by Permuth et al., reportedly distinguished indolent (benign) intraductal papillary mucinous neoplasms (IPMNs) from aggressive (malignant) IPMNs [58]. This 8 lncRNA-signature reportedly had greater accuracy than standard clinical and radiological features. It was further improved when combined with plasma miRNA data and quantitative radiomic imaging.

While early studies suggest that the analysis of circulating lncRNA levels may contribute to the evaluation of disease progression, more investigations focusing on blood-based lncRNAs are needed to truly appreciate the prognosis power of circulating lncRNAs. The best diagnostic/prognostic performance may actually emerge from the integration of several analytic methods that combine circulating lncRNA data, miRNA data, clinical data, quantitative imaging features [58] and/or conventional glycoprotein antigens such as carcinoembryonic antigen (CEA) [51] or prostate-specific antigen (PSA) [14].

**Circulating lncRNAs as potential therapeutic agents/targets for cancer treatment**

Circulating lncRNAs should not be considered only as passive biomedical tools that solely enable the detection and monitoring of various diseases. They may also constitute effective therapeutic agents and/or targets in innovative strategies that could treat various types of cancers including colorectal cancer and renal cell carcinoma [26, 62-64]. Indeed, lncRNAs have been shown to trigger or contribute to tumorigenesis notably by interfering with tumor-suppressive signaling pathways or acting as oncogenic stimuli [65-69]. In a Genome-wide analysis of the human p53 transcriptional network, Sanchez et al. notably revealed the existence of a lncRNA tumour suppressor signature [70]. GAS5, CCND1, LET, PTENP1 and lincRNA-p21 have been described as tumor suppressors [27, 62, 71-74], while MALAT-1, PANDAR, HOTAIR, H19, PVT1, GIHCG and ANRIL have been characterized as oncogenic lncRNAs [27, 62, 75-77]. At the molecular level, lncRNAs can promote tumorigenesis by acting as chromatin structure regulators that modify gene expression [78], scaffolds for oncogenic RNA-binding proteins [79] or RNA sponges for oncosuppressor microRNAs [80, 81]. For instance, lncRNA HOTTIP (HOXA transcript at the distal tip) was shown to act as a sponge for the tumor-suppressive microRNA miR-615-3p and dysregulation of HOTTIP expression was shown to alter levels of miR-615-3p and its target IGF-2, promoting the formation of RCC tumors [81]. Many more mechanisms have been described and continue to be discovered. Through various pathways, dysregulation of lncRNAs levels eventually promotes cancer cell proliferation, migration, invasion and/or metastasis [81-84]. Therefore, lncRNAs do constitute legitimate therapeutic targets. However, most mechanistic studies have been done on cancer tissues or cells, so it is still unclear if targeting lncRNAs in blood would be sufficient to treat tumors located deep inside layers of tissues. A more fundamental question may be to determine whether circulating lncRNAs can actually penetrate cells and tissues. Nucleic acids are usually unable to cross the hydrophobic cellular plasma membrane due to their large size and negative charges carried by the phosphate groups of nucleotides. In vitro DNA transfection is usually achieved by using specific carriers such as lipofectamine. Answers may come from reports indicating that circulating lncRNAs are, at least for a part, transported in the blood via extracellular vesicles such as exosomes [19]. It has even been reported that 3.36 % of the total exosomal RNA content is represented by lncRNAs [85]. Circulating exosomes are lipid-based extracellular vesicles that promote the transport of various biomolecules across long distances within the human body. Microvesicles and exosomes have notably been characterized as potent messengers that enable cancer cells to communicate with each other (autocrine messengers) and also with non-cancerous cells (paracrine and endocrine messengers [86]. Because of their lipidic structure, exosomes can fuse with the plasma membrane of a targeted cell and release their content inside it, including lncRNAs. It is thus conceivable that exosome-borne lincRNAs may be used by cancer cells to spread within the human body. Therefore, circulating lincRNAs may constitute bonafide therapeutic targets as much as tissue lncRNAs do (Figure 1). Besides exosomes, some circulating lncRNAs may be transported as complexes with circulatory proteins such as Argonaute (Ago) or nucleophosmin 1 (NPM1) similar to circulating miRNAs [87, 88]. Others may be transported in blood without any binding partner or specific protective structure. These lncRNAs may constitute the easiest targets for lncRNA-interfering cancer therapy. While the circulatory system is devoid of cellular machinery that degrades RNA-RNA and RNA-DNA hybrids, targeting lncRNAs using ASOs (RNAseH-dependent antisense oligonucleotide) can effectively produce significant antitumoral effects in vivo. Arun et al have notably shown that the systemic knockdown of Malat-1 by subcutaneous injections of ASOs in an MMTV-PyMT mouse mammary carcinoma model resulted in slower tumor growth and a reduction in metastasis [89].

Other studies have highlighted the existence of lncRNAs that are downregulated in cancer tissues [90] and the circulation of cancer patients [42]. Such downregulated lincRNAs may be oncosuppressor lncRNAs of which expression is dysregulated during tumorigenesis. The ectopic delivery of synthetic or purified oncosuppressor lncRNAs may constitute a promising therapeutic strategy in the future (Figure 1). These therapeutic oncosuppressor lncRNAs may be administrated as an exosome-based formula which could possibly treat primary and secondary tumors as it spreads throughout the body via the circulatory system.

If some circulating lncRNAs are indeed shown to have oncosuppressive properties in vivo, they may also be uptaken prior to cancer formation for cancer prevention purposes, similar to anti-oxidants (Figure 1).

**Cancer-specific, multi-cancer and pan-cancer circulating lncRNA biomarkers and therapeutic targets**

A significant number of circulating lncRNAs have been reported to be associated with only one cancer type so far (Table 1). While this could be due to a lack of studies on these lncRNAs in other cancer types, it could also imply that certain blood-based lncRNAs may really be specific to a unique type of cancer only, which has significant translational applications especially in cancer screening since the detection of abnormal levels of such lncRNAs in the circulation would not only be indicative of a cancer diagnosis but also pinpoint with accuracy the organ affected by the tumor. More studies need to be undertaken to evaluate the plausibility of these two scenarios. Interestingly, the integrated analysis of the most reported circulating lncRNAs and their specific association with certain cancers seems to reveal a pattern where some circulating lncRNAs are apparently able to reflect multiple cancers especially in organs that are close anatomically and/or embryologically (Figure 2A, lncRNAs in white letters). For instance, circulating LINC00152, HULC and UCA1 have been associated with gastric and liver cancer, two organs that are in close proximity within the upper abdomen and which both originate from the foregut of the embryonic endoderm [19, 22, 38-40, 91]. Lung and esophagus which are located in the thorax and share common embryological origins (before they split apart during development) also show a similar circulating lncRNA - SPRY4-IT1 - upon tumorigenesis [52, 53]. Circulating HOTAIR has been detected in the blood of patients with cancers of the uterus and colon/rectum, organs that are located in the pelvis and sometimes fused in congenital diseases such as persistent cloaca [37, 92]. Levels of circulating lncRNAs PVT-1 and PANDA reportedly reflect tumorigenesis or malignancy in the kidney and pancreas, two organs that are in close proximity and often grafted together [57, 58]. Circulating PVT-1 also reflects tumor formation in the liver, an organ close anatomically and embryologically to the pancreas [54]. The fact that cancers from the same anatomical region or embryological origin display a similar circulating lncRNA molecular signature is consistent with the findings from an integrative study published in 2018 that analyzed the complete set of tumors in The Cancer Genome Atlas (TCGA), consisting of approximately 10,000 specimens and representing 33 cancer types [93]. In this study, the authors performed molecular clustering based on RNA expression levels and other key features and concluded that clustering is primarily organized by histology, tissue type, or anatomic origin [93]. Moreover, the embryological origin of human tumors has been largely discussed and is notably supported by evidence suggesting that adult somatic cells retain an embryonic program that can be reactivated in certain pathological conditions promoting the dedifferentiation into stem cells and eventually tumorigenesis [94]. In addition, machine learning has enabled the identification of key stemness features that are associated with oncogenic dedifferentiation [95] while embryonic stem cell-like gene expression signatures have been identified in human tumors [96-98]. Because of their involvement in both tumorigenesis and development, several genes including some coding for lncRNAs have been referred to as "oncofetal" [99]. They are reportedly upregulated in the embryo and downregulated in adults [100]. However, in some cancers, these oncofetal lncRNAs may be re-expressed contributing to tumorigenesis and malignancy [101]. In this context, cancer may arise due to loss of cellular differentiation and gain of pluri- or multi-potency with the high proliferative potential characteristic of stem cells [102]. This concept notably led to the characterization of cancer stem cells. In fact, it is believed that, as somatic cells from different organs of the same anatomic region dedifferentiate into cancer stem cells, they may indirectly try to recreate the same embryonic organ that was originally responsible for their formation during embryogenesis (which they share in common). Based on this cumulative information, it is perhaps not surprising to observe similar patterns of blood lncRNA levels in cancers with the same embryological or anatomical origin as shown in Figure 2A and 2B. However, there are some exceptions and circulating lincRNAs may not necessarily change upon tumorigenesis according to organ location or its embryological origin (e.g. endoderm, mesoderm, ectoderm). For instance, circulating lncRNAs associated with cancer from organs related to reproduction (e.g. prostate, breast) may not follow such an anatomic/embryonic pattern as sexual organs are usually not developed during embryogenesis. Although, in healthy adults, sexual organs appear to be the main sources of some of the most widely reported cancer-associated lncRNAs such as PVT1 and MALAT1 that are mostly expressed in the ovaries of healthy women, while PTENP1 is largely expressed in the testis of healthy men (Figure 2C). Those lncRNAs mostly remain poorly expressed in other tissues of healthy individuals. The fact that many of these lncRNAs are suppressed in most adult tissues but remain extensively expressed in sexual organs (either ovaries or testis, exclusively) suggests the likely involvement of so-called "genomic imprinting". It essentially consists in the reprogramming of the epigenetic make-up of certain key genes according to the sex of the individual during gametogenesis, which results in the fetus in a parent-of-origin type of gene expression with transcription occurring only on one allele while being suppressed on the other (notably through DNA methylation and histone modification). *H19* for instance is an imprinted gene that is known to be transcribed exclusively from the maternal allele and silenced on the paternal allele [103]. *H19* is in fact the first imprinted lncRNA-encoding gene ever identified [100] and its product, the lncRNA H19 (H19 Imprinted Maternally Expressed Transcript), has since been the object of numerous studies to understand its implications in health and disease. H19 lncRNA has notably been reported to play critical roles in both developments [104-106] and tumorigenesis [107-114] and therefore legitimately belongs to the class of oncofetal lncRNAs [99, 115, 116]. A major mechanism by which imprinted lncRNAs such as H19 induce or contribute to tumorigenesis likely involves a still poorly understood event known as "loss-of-imprinting" or LOI that abnormally restores gene expression on both alleles (i.e. "biallelic expression") in adult somatic cells potentially promoting cancer formation. The reasons for sporadic LOI are not fully understood but likely involve the partial or complete loss of the imprinted epigenetic code of certain key regulatory regions within the DNA sequence notably due to major changes in methylation patterns (e.g. hypomethylation or hypermethylation) that can reportedly be induced by exposure to cigarette smoke for instance. This may affect the ability to recruit insulating proteins such as CTCF resulting in changes in chromatic structure including de-condensation potentially promoting gene expression on the allele that should otherwise be suppressed, Eventually, it is undeniably clear that circulating imprinted lncRNAs that are expressed during development and which reflect, in adults, tumors from organs with a same embryonic origin could constitute potential "oncofetal imprinted lncRNA biomarkers" as well as promising therapeutic targets. These embryo-derived lncRNAs do represent promising multi-cancer biomarkers that would not only enable the detection of various types of cancers but also determine the likely location of the tumor in the adult body as well as the organ(s) affected by tumorigenesis. Embryo-related biomarkers such as the carcinoembryonic antigen (CEA) are already in use for the diagnosis of many cancers.

The existence of potential pan-cancer circulating lncRNA biomarkers has also been investigated including by our lab. Indeed, in a leading study based on the rigorous and systematic statistical analysis of gene expression profiles of twelve different cancer types extracted from multiple publicly available databases, our lab identified 6 promising pan-cancer lincRNA biomarkers subsequently termed "PCAN" lincRNAs that are systematically dysregulated in cancer [90]. Active efforts are currently undertaken to explore the full potential of these PCAN lincRNAs by extending the study to cancers beyond the original 12 cancer types. Upon validation in blood-based samples, this panel of PCAN biomarkers could potentially constitute the first set of circulating lincRNAs capable of detecting any kind of cancer in the human body. Further investigations would also be required to better understand the molecular mechanisms associated with the upregulation of these PCAN lncRNAs in cancer and to assess whether they could constitute potential pan-cancer therapeutic targets as well as imprinted oncofetal genes similar to *H19*.

**Circulating lncRNAs and association with RNA-binding proteins**

While RNA-binding proteins may not interact with circulating lncRNAs once they reach the bloodstream, they may bind lncRNAs inside the tumor cells prior to secretion and may actively contribute to the tumorigenic process. Indeed, many RNA-binding proteins that interact with lncRNAs have also been characterized as oncofetal [117, 118]. This suggests that lncRNA-related tumorigenesis is likely the result of a complex and diversified molecular mechanism that involves the upregulation of several oncofetal genes, including genes coding for oncofetal lncRNAs and oncofetal lncRNA-binding proteins. Investigators can find information of lncRNA-binding partners by screening databases such as lncRNome, lncRNAMap, starBase V2.0 and UCSC genome browser [119, 120]. The systematic analysis of these databases actually revealed a common set of proteins that consistently interacts with the most reported cancer-related lncRNAs (Figure 3A). Most of these proteins are associated with cancer formation upon dysregulation, especially IGF2BP3 [121, 122], FUS [123, 124] and eIF4A3 [125]. This suggests the likely existence of a pan-lincRNA core protein interactome that may, by itself, be sufficient to promote tumorigenesis. However, some of these proteins appear to be more frequently involved in lncRNA interactions than others and may play a more central role in cancer formation. For instance, eIF4A3 was found to interact with 9 of out 10 lncRNAs in the lncRNA panel reported here (Fig.3A, 3B), while FUS was recruited by 8 out of 10 lncRNAs. Therefore, eIF4A3 and FUS may constitute key lncRNA-binding proteins that could be part of a pan-cancer molecular mechanism that mediates the tumorigenic properties of most oncogenic lncRNAs and/or generally promotes lncRNA secretion into the systemic circulation from the tumor site. Thus, eIF4A3 and FUS may represent major pan-cancer therapeutic targets. While other RNA-binding proteins appear to be less frequently recruited by cancer-related lncRNAs, they may still exert pan-tumorigenic properties since all RNA-binding proteins reported here in Fig.3A are part of a very same multimeric protein complex based on data from an extensive search of protein-protein interactions using BioGRID database (Figure 3C). Interestingly, eIF4A3 and FUS showed the highest ability to interact with other RNA-binding proteins (respectively binding 4 and 5 other protein partners within the complex), which may explain why they are often associated with lncRNAs since the more lncRNA-binding proteins they bind, the more lncRNAs they collect. Overall, it is clear that lncRNAs and their interacting partners will constitute innovative therapeutic targets and/or agents in future cancer therapy strategies.

**Guidelines for the study of blood-based lncRNAs as biomarkers for early cancer diagnosis**

*Source of circulating RNAs*

Prior to proceeding with RNA extraction from blood, investigators should choose an appropriate and reliable source of lncRNAs. While whole blood has been successfully used in circulating lncRNA studies [42], it is usually not recommended for accurate quantification of circulating RNAs due to variability associated with red and white blood cells [126]. Indeed, the level of circulating white blood cells (and thus circulating RNAs) is likely to change significantly if patients studied are experiencing chronic or acute inflammation which may or may not be related to the disease investigated [127, 128]. Inflammation and immune responses may occur for a variety of reasons including tissue injuries, infections, allergies, auto-immune responses and even some medications. Therefore, it is recommended to not use whole blood as a source of circulating lncRNAs and to exclude patients experiencing inflammatory processes from studies especially those involving cancer patients (Table 2).

Cell-free blood-derived samples such as plasma (the blood fraction obtained with anti-coagulants) and serum (the blood fraction obtained after coagulation) have been described as more reliable sources of circulating lncRNAs (Table 2). Serum and plasma have been largely used in studies comparing circulating lncRNA levels in cancer patients and healthy subjects (Table 1). Plasma and serum appear to be relatively similar in terms of RNA content but the lack of comparative studies prevents any clear conclusion. LncRNA levels in serum or plasma may change dramatically upon disease development especially during tumorigenesis. Indeed, pathological processes associated with cancer progressions such as tissue injury and cell death are likely to promote the release of cellular content - including lncRNAs - into the circulation. Levels of circulating RNAs may also vary within the same group of individuals (e.g. healthy volunteers) due to internal factors such as patient hydration level or diet [129, 130]. Age, gender and race are also factors that are likely to affect the levels of circulating lncRNAs (Table 2). Copy number variations (CNVs) and single nucleotide polymorphisms (SNPs) have notably been proposed as possible sources of variations in levels of circulating lncRNAs. Consequently, investigators usually collect relevant patient information and subsequently compare cancer patients and healthy subjects with similar records. It is thus important to note that equal volumes of plasma samples may not contain the same concentration of total RNAs. Inconsistent serum or plasma preparation across samples may also add another level of variability in the final RNA content especially if hemolysis could not be avoided. To account for hemolysis, Permuth et al. visually inspected their samples and measured absorbance at three different wavelengths (414nm, 541nm and 576nm). An absorbance exceeding 0.2 for any of these wavelengths indicated hemolyzed samples [58].

They further assessed for blood-cell contaminants by measuring levels of transcripts from MB, NGB and CYGB genes (for erythrocytes) as well as APOE, CD68, CD2 and CD3 (for white blood cells).

*RNA extraction methods*

Extraction of lncRNAs from serum or plasma samples also represents a crucial step in the quantification of circulating lncRNAs. Different RNA extraction methods and protocols are currently available but very few matches exactly the investigators' needs, especially regarding the specific extraction of lncRNAs from plasma or serum samples. Investigators that wish to use column-based kits should be aware that most commercially available kits are optimized for non-liquid samples such as cells or tissues, and not for plasma or serum. Some kits such as the miRNeasy Serum/ Plasma kit do allow RNA extraction from serum and plasma, but it is mostly designed for purification of microRNAs (miRNAs) and other small RNAs. Since lncRNAs are naturally scarce in circulation, investigators may wish to use large volumes of plasma or serum to increase the RNA yield upon extraction. However, most kits are provided with columns of limited size which may introduce variability in RNA yields as investigators often have to perform successive column-based purifications with small volumes of the same sample. If different kit formats are available (for instance mini, midi and maxi), investigators should proceed with the kit that is the most suitable for their study based on the volume of samples that is available to them. It is important to note that if the volume of the original plasma sample is too small, the RNA yield might be too low for RNA quantification and qPCR detection. Despite the relatively low RNA yield generated with blood-based samples, most kits provide RNA samples of high purity due to the solid phase extraction method using glass fiber filters for instance as well as multiple washing steps. Improvements to the RNA extraction method may come from the addition of an organic extraction based on liquid phase separation using phenol/chloroform (Table 2). For instance, the mirVana kit which combines both solid phase (filter) and liquid phase (chloroform) RNA extraction has been largely used in cancer studies focusing on circulating lncRNAs [17, 28, 37]. This kit appears to be popular among investigators because it allows not only total RNA extraction from liquid samples (plasma/serum) but also purification of small RNAs and long RNAs such as lncRNAs.

*Reverse Transcription and qPCR*

The RNAs extracted from plasma or serum samples subsequently undergo reverse transcription (RT) to enable single-strand cDNA synthesis before qPCR amplification. In the literature, there is no real consensus on what the best method is to achieve RT and qPCR for lncRNAs extracted from plasma or serum samples. Especially, there is clear clack of unanimous standards for the reliable normalization and quantification of circulating RNA levels across biological samples. Investigators basically have the choice to use equal amounts of total RNAs or equal volumes of RNA extracts as inputs for reverse transcription. Considering the variations in RNA concentration across samples, using equal quantities of total RNAs may appear as a more accurate method for the comparative analysis of circulating lincRNAs between case and control samples. However, the quantity of total RNAs obtained after RNA extraction from plasma or serum samples is usually very low and is often undetectable using spectrophotometers such as NanoDrop. Using equal RNA quantities also implies the adjustment of all samples to the sample with the lowest RNA concentration which may reduce the output of the subsequent qPCR reaction. For the analysis of lncRNAs in body fluids that are naturally poor in RNAs such as plasma and serum, it might thus be more convenient to use equal volumes of RNA extracts for the reverse transcription step (Table 2). In this context, the investigators may be able to use the maximum template volume available for an RT reaction for optimal cDNA synthesis (for example 14µl of RNA extract for an RT reaction of 20µl that includes non-compressible 6µl of enzyme and buffer). Consequently, the Ct values obtained in qPCR may have a lower – better - range providing more reliable data (high Ct values around 40 for instance are usually considered not reliable).

If investigators choose to proceed with equal volumes of RNA extracts, normalization using reference genes as internal controls becomes indispensable for the analysis of qPCR data by relative quantification (delta delta Ct / ΔΔCt method). The literature indicates that various genes can be used as references including traditional housekeeping genes, but there is currently no common agreement on what the best reference genes are. They appear to be case-sensitive and have to be tested by the investigators for each disease studied. Some of the most frequently used in cancer studies are GAPDH RNA, β-actin RNA and 18S RNA (Table 1 , 2). These RNAs are usually present in high quantities in plasma and serum samples which makes them more reliable than other RNAs that are poorly present in the blood. In theory, an ideal reference gene should show no differences in quantities (reflected by Ct values) between control and cancer samples. In practice, it is rarely the case. Variations are inevitable and rather natural considering the various physiopathological processes associated with tumorigenesis including inflammation, tissue damage and cell death that are likely to induce the release of nucleic acids into the circulation (including potential reference RNAs) [126, 128]. Ultimately, it is more a matter of determining whether these variations are statistically significant or not [35]. Even then, it is important to remember that an apparently insignificant difference in Ct values of only 1 usually corresponds to a fold change of 2 (if qPCR efficiency is maximal). Usually, investigators test a panel of different potential reference genes and select the least variable candidate by using algorithms such as NormFinder, RefFinder or Genorm [32, 35, 37]. Variability in Ct values of potential reference gene candidates may arise within the same group of individuals (e.g. healthy subjects) and/or between groups of individuals (cancer patients and healthy subjects). Variations between groups may be acceptable as long as variation within the same group remains minimal allowing reliable relative qPCR quantification using the ΔΔCT method. For instance, Dong et al., selected β-actin as the reference gene for their analysis of circulating lncRNAs CUDR, LSINCT-5 and PTENP1 in sera of gastric cancer patients despite small variations in β-actin Ct values between control and cancer samples [32]. Thus, small variations in Ct values of a potential candidate reference gene may not constitute an automatic reason for withdrawing the reference gene as a potential internal control for relative qPCR quantification. The selection of the best reference gene may sometimes be time-consuming and low throughput (one or two genes will be selected among many). It may also require the commitment of precious and limited resources such as a significant number of biological samples (cancer samples and matching controls) in order to produce statistically significant data.

Because of their poor abundance in the bloodstream or recurrent variability, several genes should not be used as internal controls when investigating circulating lncRNAs. For instance, RPLPO which is a reliable reference gene for tissue sample analysis is not recommended for the quantification of blood-based biomarkers [35, 131] (Table 2). Other genes like HPRT and GUSB also have to be avoided in circulating lncRNA studies due to very low abundance in normal human serum [32].

The use of exogenous spike-in controls added during the RNA extraction process may also be used to account for any variations introduced during the RNA extraction process and subsequent experimentations. However, spike-in controls do not reflect inherent variations in RNA concentrations in blood-derived samples prior to RNA extraction [132]. For instance, normalization with spike-in controls does not account for RNA variations across plasma samples that are induced by different patient hydration levels, diets or uncontrolled hemolysis during sample preparation from whole blood (Table 2). On the contrary, normalizing with spike-in controls may sometimes be detrimental to the quantification of circulating lncRNAs. Indeed, any difference in lncRNAs noticed between plasma samples from cases and controls might be wrongly attributed to the disease and not to the simple variation in sample RNA concentration. Therefore, we suggest using, on top of spike-in controls, blood-based internal controls as reference genes to better account for variations in circulating lncRNA levels between cancer and control samples, as reported in the literature (Table 1). One or more reference genes may be used to accurately evaluate changes in lncRNAs levels between case and control samples using the delta delta Ct ($\Delta\Delta Ct$) method. If no appropriate reference gene is identified, investigators may want to evaluate lncRNA levels by absolute qPCR quantification using standard curves made with reference standards [30].

*Next-generation sequencing-based technologies*

Next-generation sequencing (NGS) technologies have been widely applied to the cancer field in the past decades. There are two major paradigms in next-generation sequencing technology: short-read sequencing (35–700 bp) [133] and long-read sequencing (>5000bp) [134]. Short-read sequencing approaches provide lower-cost, higher-accuracy data that are useful for population-level research and clinical variant discovery [133], while long-read approaches provide read lengths that are well suited for de novo genome assembly applications and full-length isoform sequencing [134]. In practice, short-read sequencing such as RNA sequencing (RNA-Seq) technology is usually used for cancer early diagnosis [135].

The lncRNAs' concentration is about one order of magnitude lower than that of mRNA and represents only 1–2% of polyadenylated RNAs in a cell. It is necessary to first enhance their concentration by building lncRNA-specific cDNA library using oligonucleotide capture technology [136]. With complementary oligonucleotide probes, this technology increases the concentration of lncRNA sequences by at least 25% [136]. The first step of the oligonucleotide capture technology is the design of complementary oligonucleotide probes, which are targeted to lncRNA transcripts of interest. The careful design and forethought at this step are important for the subsequent success of the lncRNA specific cDNA library. Targeting only part of a lncRNA is usually sufficient to guide and capture the full-length lncRNA. The optimization of probe sequences is quite similar to that for microarray technologies [137]. The following steps constitute the preparation of cDNA library, hybridization of library to oligonucleotide probes, washing and removal of non-targeted cDNA and elution of targeted cDNA for sequencing. While PCR amplification is required both for the cDNA library preparation before hybridization and for the amplification of the captured lncRNA-specific cDNA library before sequencing, it also adds the risk of unwanted amplification artifacts or biases. To minimize the amplification artifacts, it is better to reduce the number of amplification cycles as much as possible. It's recommended to evaluate the fold enrichment and capture performance using qPCR before sequencing. Performing a control DNA capture using matched genomic DNA (gDNA) will also be helpful to identify anomalous probe hybridization artifacts and normalize lncRNA expression.

The computational preprocessing of lncRNA sequencing data becomes more or less routine over the years, starting from raw NGS fastq data. The first step is to use QC tools to filter of low-quality reads from raw data using pre-processing tools, such as FastQC [138], PRINSEQ [139] and AfterQC [140]. The next step is to align processed clean reads to a noncoding transcriptome reference, such as LNCipedia [141], using tools like HISAT2 [142] or BowTie2 [143]. Then the mapped reads are assembled to lncRNAs using tools like StringTie2 [144]. The quantity of lncRNAs is to be normalized by the sequencing depth and transcript length using RPKM (Reads Per Kilobase Million) or FPKM (Fragments Per Kilobase Million) or TPM (Transcripts Per Kilobase Million) method [145]. The differentially expressed lncRNAs between cancer and normal are further analyzed by specialized tools like DESeq2 [146], or edgeR [147]. Besides using differential expression p-values, risk score analysis can be used to determine the association between cancer and each top differentially expressed lncRNAs [148]. By associating lncRNAs with the nearest protein coding genes or some other approaches, they can be further annotated by gene ontology (GO) [149] and Kyoto Encyclopedia of Genes and Genomes (KEGG) pathway [150] analyses to explore their biological functions. Finally, with sufficient number of samples, classification models can be used to identify potential lncRNA biomarkers, with individual training, testing and valiation datasets [151, 152]. Metrics such as area under the curve (AUC) for Receiver Operating Characteristic (ROC) are then used to measure the effectiveness of the lncRNAs in predicting cancer [90].

**Discussion and future perspectives**

Circulating lncRNAs have been shown to constitute reliable biomarkers for both cancer diagnosis and prognosis. They have also been suggested as potential therapeutic targets. While diagnostic performance can be improved by combining multiple lncRNAs, it is important to note that the "specificity" determined in the reported studies refers to the comparative analysis of samples from healthy volunteers and patients with specific cancer. In this particular context, "specificity" does not describe the ability to distinguish a certain cancer type from other cancers. This is particularly relevant since several circulating lncRNAs have been proposed as potential biomarkers for a large variety of different cancers. For instance, MALAT-1 could be used to diagnose prostate cancer [28] and non-small-cell lung cancer [35, 42]. Similarly, HOTAIR has the potential to detect both colorectal [37] and cervical cancer [92]. LINC00152 could lead to the diagnosis of both hepatocellular carcinoma [38] and gastric cancer [19]. LncRNA GIHCG has been shown to be involved in the pathogenesis of many types of different cancers including liver, cervical, gastric, renal and colorectal cancer for which it may constitute a promising biomarker [26, 36, 77, 153-155]. PVT1 has been reported as a potential circulating biomarker (alone or in combination with other lncRNAs) for at least five different types of cancers including RCC (kidney), IPMN (pancreas), HCC (liver), MLN (skin), and CVC (cervix) [54, 57, 156, 157]. UCA1 constitutes another lncRNA with significant multi-cancer diagnostic potential since it has been reported to effectively detect (alone or in combination with other lncRNAs) at least four distinct cancers such as HCC (liver), GC (stomach), BC (bladder) and CRC (colon) [32, 39, 40, 91, 158]. The increasing number of studies on circulating lincRNAs may eventually indicate that all circulating lncRNAs reflect more than one cancer and that there is no unique biomarker for each cancer type or subtype. It has especially been suggested that changes in lncRNA level in the circulation of cancer patients could be due to a general physio-pathological response from the body to the presence of tumors and not due to direct secretions from the tumors themselves [132]. This represents a strong argument as significant levels of lncRNAs have been detected in the blood of cancer-free healthy subjects. This would also explain why there is sometimes a lack of correlation between circulating lncRNA levels and cancer tissue lncRNA levels. Thus, circulating lncRNAs may actually reflect the presence of tumors in general. In this context, it is likely that in the near future pan-cancer circulating biomarkers could be identified. While the study of circulating lncRNAs is still at an early stage, the growing interest increases the chance of discovering blood-based biomarkers that will allow the early and accurate detection of cancer before it becomes uncurable.

**Author Contributions**

LXG envisioned the project. CB wrote the manuscript with the help of BH and LXG. All authors have read, revised and approved the final manuscript.

**Acknowledgements**

This research was supported by grants K01ES025434 awarded by NIEHS through funds provided by the trans-NIH Big Data to Knowledge (BD2K) initiative (www.bd2k.nih.gov), P20 COBRE

**Conflict of Interest**

The authors declare that there is no conflict of interest.

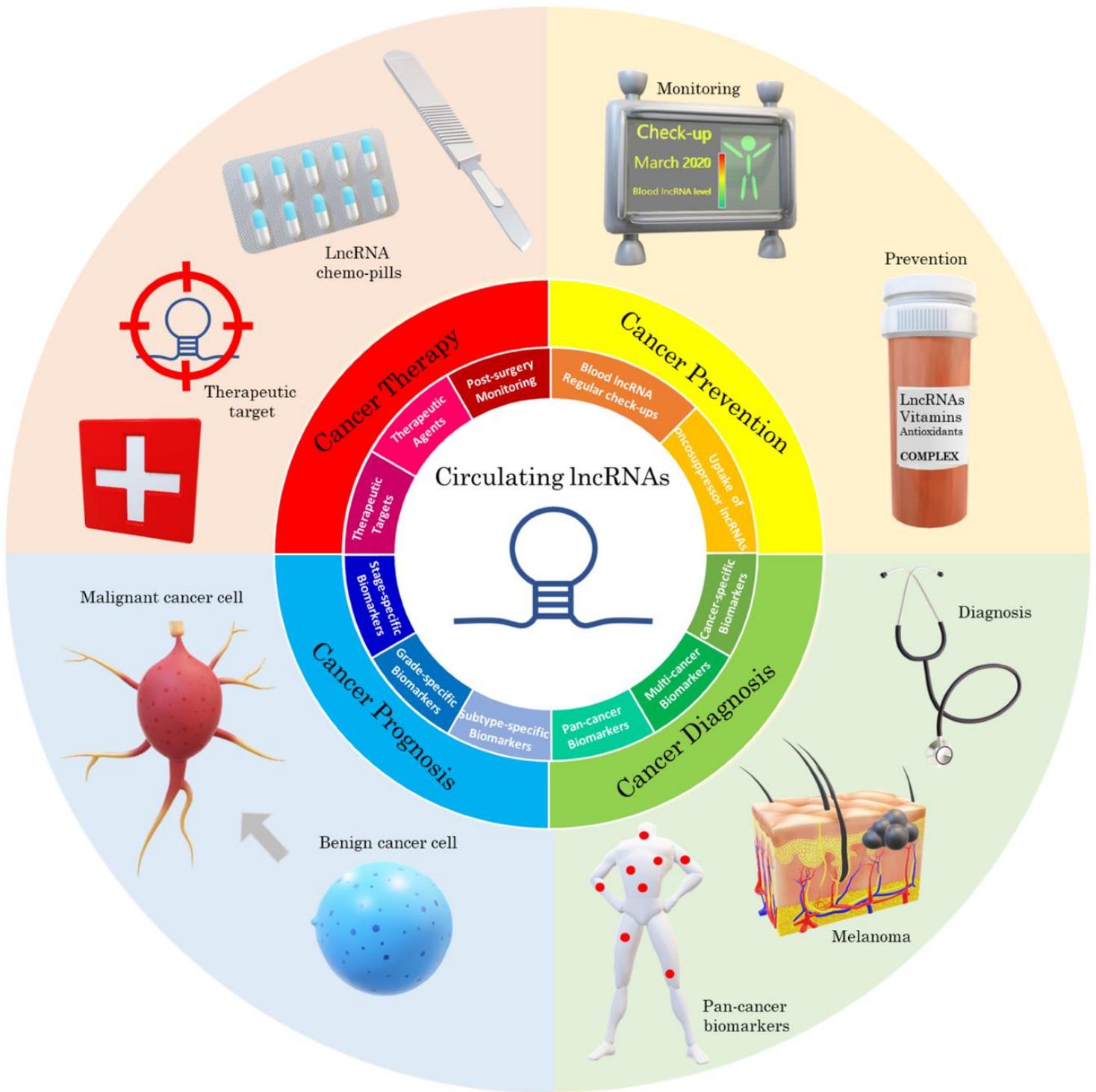

**Figure 1**. **Diagram summarizing the full panel of possible clinical applications that can be derived from the analysis of blood-based lncRNA**s. Information indicated includes four main domains of applications (cancer prevention, cancer diagnosis, cancer prognosis, cancer treatment) and smaller subdomains referring to the domain of the same color.

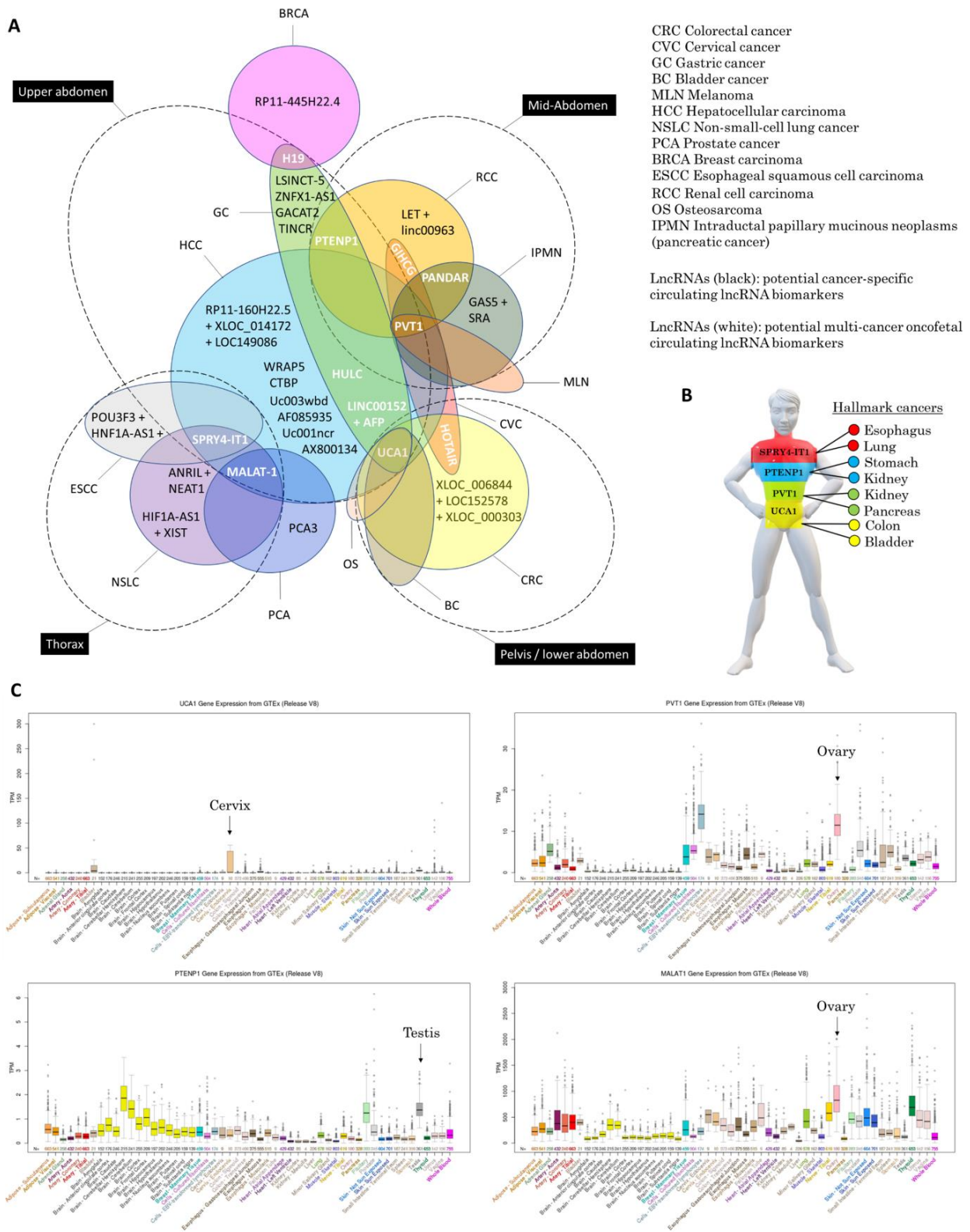

**Figure 2. Cancer-specific and multicancer blood-derived lncRNA biomarkers.** A. Diagram showing circulating lncRNAs reported in the literature regrouped by cancer type. Some lncRNAs (in black letters) are cancer-specific. Other circulating lncRNAs (in white letters) such as MALAT1, SPRY4-IT1, PVT1, UCA1 and LINC00152 reflect tumorigenesis in multiple organs. B. Simplified cartoon representing the specificity of certain circulating lncRNAs towards cancers of organs located in particular anatomic segments of the human body. C. Gene tissue expression of some of the most widely reported circulating lncRNAs with high multi-cancer diagnosis potential (GTEx, obtained from UCSC genome browser).

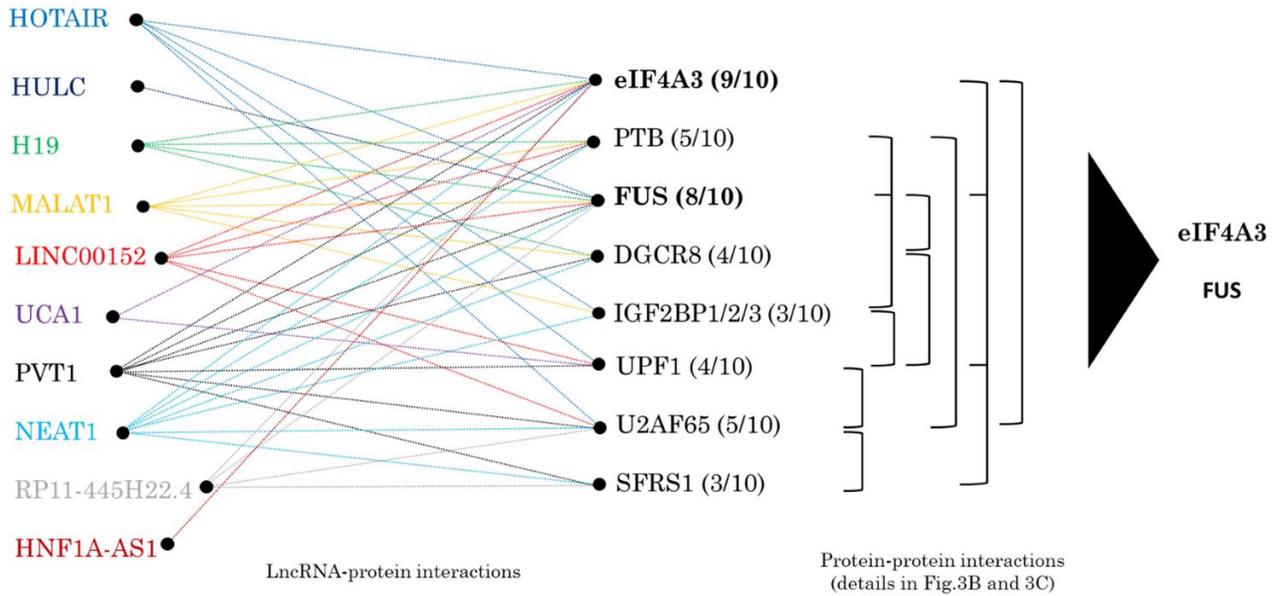

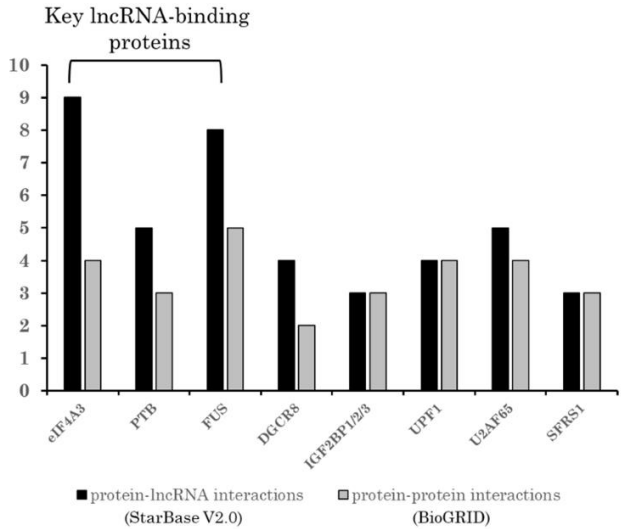
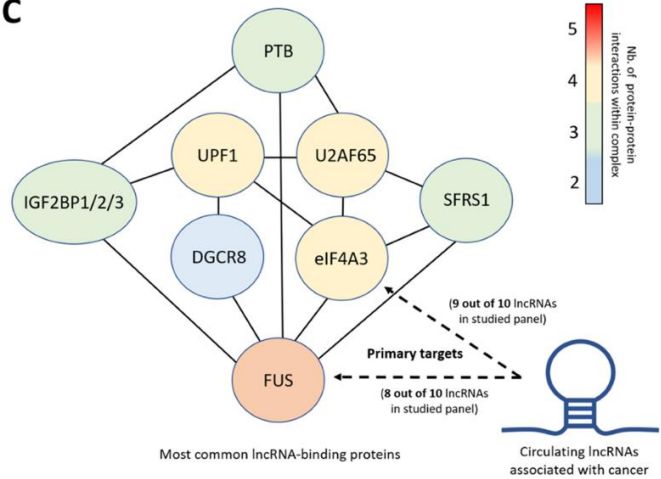

**Figure 3. Circulating lincRNAs and a common set of protein partners**. A. Data extracted from starBase V2.0 and lncRNome databases reporting lncRNA-protein interactions occurring in tissues. Indicated lncRNAs share the same set of interacting proteins that are also known to be involved in tumorigenesis. These main proteins may constitute an oncogenic pan-lncRNA core protein interactome. Displayed protein-protein interactions are based on data from BioGRID database. B. Graph bars representing the number of interactions with lncRNAs and proteins for each RNA-binding protein shown in (A). C. Putative pan-cancer multimeric RNA-binding protein complex showing the different interactions between the proteins that are the most commonly recruited by cancer-related lncRNAs as shown in (A).

| LncRNA | Cancer type | Source | Sensitivity (%) | Specificity (%) | AUC / QUADAS | Normalization | Reference |
|---|---|---|---|---|---|---|---|
| MALAT-1 | Non-small-cell lung cancer | Blood cells | 56 | 96 | AUC 0.79 | GAPDH | Weber et al., 2013 |
| | Non-small-cell lung cancer | Whole blood | N.A. | N.A. | AUC 0.718 | GAPDH | Guo et al., 2015 |
| | Prostate cancer | Plasma | 58.6 | 84.8 | AUC 0.836 | Standard curve | Ren et al., 2013 |
| | Hepatocellular carcinoma | Plasma | 51.1 | 89.3 | AUC 0.66 | MALAT-1 | Konishi et al., 2016 |
| LINC00152 | Gastric cancer | Plasma | 48.1 | 85.2 | AUC 0.675 | GAPDH | Li et al., 2015b |
| | Hepatocellular carcinoma | Plasma | N.A. | N.A. | AUC 0.85 | 5S | Li et al., 2015a |
| | Hepatocellular carcinoma | Serum | 78.3 | 89.2 | AUC 0.877 | GAPDH | Huang et al., 2020 |
| UCA1 | Hepatocellular carcinoma | Serum | 92.7 | 82.1 | AUC 0.861 | GAPDH | Kamel et al., 2016 |
| | Hepatocellular carcinoma | Serum | 91.4 | 88.6 | QUADAS 11 | ß-actin | El-Tawdi et al., 2016 |
| | Colorectal cancer | Plasma | N.A. | N.A. | N.A. | Cel-miR-39 | Tao et al., 2015 |
| | Gastric cancer | Plasma | N.A. | N.A. | AUC 0.928 | GAPDH | Gao et al., 2015 |
| | Osteosarcoma | Serum | N.A. | N.A. | AUC 0.831 | GAPDH | Wen et al., 2017 |
| H19 | Gastric cancer | Plasma | 74 | 58 | AUC 0.64 | LncRNA levels | Arita et al., 2013 |
| | Gastric cancer | Plasma | 82.9 | 72.9 | AUC 0.838 | Standard curve | Zhou et al. 2015 |
| | Gastric cancer | Plasma | 68.75 | 56.67 | AUC 0.724 | GAPDH | Hashad et al., 2016 |
| | Breast cancer | Plasma | 56.7 | 86.7 | AUC 0.81 | ß-actin | Zhang et al., 2016 |
| PVT1 | Cervical cancer | Serum | 71.6 | 98.8 | AUC 0.932 | GAPDH | Yang et al., 2016 |
| | Melanoma | Serum | 94.12 | 85.11 | AUC 0.938 | GAPDH | Chen et al., 2017 |
| WRAP53 | Hepatocellular carcinoma | Serum | 85.4 | 82.1 | AUC 0.896 | GAPDH | Kamel et al., 2016 |
| HULC | Hepatocellular carcinoma | Blood cells | N.A. | N.A. | N.A. | ß-actin | Panzitt et al., 2007 |
| | Hepatocellular carcinoma | Plasma | N.A. | N.A. | N.A. | GAPDH | Xie et al, 2013 |
| | Hepatocellular carcinoma | Plasma | N.A. | N.A. | AUC 0.78 | 5S | Li et al., 2015a |
| | Gastric cancer | Plasma | 58 | 80 | AUC 0.65 | GAPDH | Xian et al., 2018 |
| HOTAIR | Colorectal cancer | Blood cells | 67 | 92.5 | AUC 0.87 | PPIA | Svoboda et al., 2014 |
| | Cervical cancer | Serum | N.A. | N.A. | N.A. | GAPDH | Li et al., 2015c |
| CTBP | Hepatocellular carcinoma | Serum | 91 | 88.5 | QUADAS 11 | ß-actin | El-Tawdi et al., 2016 |
| GIHCG | Renal cell carcinoma | Serum | 87 | 84.8 | AUC 0.920 | N.A. | He et al., 2018 |
| | Cervical cancer | Serum | 88.7 | 87.5 | AUC 0.940 | ß-actin | Zhang et al., 2019 |
| PCA3 | Prostate cancer | Periph. Blood | 32 | 94 | N.A. | N.A. | Neves et al., 2013 |
| RP11-445H22.4 | Breast cancer | Serum | 92 | 74 | AUC 0.904 | U6 | Xu et al., 2015 |
| uc003wbd | Hepatocellular carcinoma | Serum | N.A. | N.A. | AUC 0.86 | ß-actin | Lu et al., 2015 |
| AF085935 | Hepatocellular carcinoma | Serum | N.A. | N.A. | AUC 0.96 | ß-actin | Lu et al., 2015 |
| GACAT2 | Gastric cancer | Plasma | 87 | 28 | AUC 0.622 | GAPDH | Tan et al., 2016 |
| SPRY4-IT1 | Hepatocellular carcinoma | Plasma | 87.3 | 50 | QUADAS 12 | 18S | Jing et al., 2016 |
| uc001ncr | Hepatocellular carcinoma | Serum | N.A. | N.A. | AUC 0.885 | GAPDH | Wang et al., 2015 |
| AX800134 | Hepatocellular carcinoma | Serum | N.A. | N.A. | AUC 0.925 | GAPDH | Wang et al., 2015 |
| ZNFX1-AS1 | Gastric cancer | Plasma | 84 | 68 | AUC 0.85 | GAPDH | Xian et al., 2018 |
| LINC00152 + AFP | Hepatocellular carcinoma | Serum | 85.3 | 83.4 | AUC 0.906 | GAPDH | Huang et al., 2020 |
| XIST + HIF1A-AS1 | Non-small-cell lung cancer | Serum | N.A. | N.A. | AUC 0.931 | GAPDH | Tantai et al., 2015 |
| PVT1 + uc002mbe.2 | Hepatocellular carcinoma | Serum | 60.5 | 90.6 | QUADAS 11 | GAPDH | Yu et al., 2016 |
| GAS5 + SRA | Pancreatic cancer (IPMN) | Plasma | 82 | 59 | AUC 0.729 | ß-actin PGK1 PPIB | Permuth et al., 2017 |
| SPRY4-IT1 + ANRIL + NEAT1 | Non-small-cell lung cancer | Plasma | 82.8 | 92.3 | AUC 0.876 | N.A. | Hu et al., 2016 |
| LINC00152 + UCA1 + AFP | Hepatocellular carcinoma | Serum | 82.9 | 88.2 | AUC 0.912 | GAPDH | Huang et al., 2020 |
| CUDR (UCA1) + LSINCT-5 + PTENP1 | Gastric cancer | Serum | 81.8 | 85.2 | AUC 0.829 | ß-actin | Dong et al., 2015 |
| SPRY4-IT1 + POU3F3 + HNF1A-AS1 | Esophageal squamous cell carcinoma | Plasma | 72.8 | 89.4 | AUC 0.842 | GAPDH | Tong et al., 2015 |
| XLOC_006844 + LOC152578 + XLOC_000303 | Colorectal cancer | Plasma | 80 | 84 | AUC 0.975 | N.A. | Shi et al., 2015 |
| RP11-160H22.5 + XLOC_014172 + LOC149086 | Hepatocellular carcinoma | Plasma | 82 | 73 | AUC 0.896 | ß-actin | Tang et al., 2015 |
| LET + PVT1 + PANDAR + PTENP1 + linc00963 | Renal cell carcinoma | serum | 67.6 | 91.4 | AUC 0.823 | ß-actin | Wu et al., 2016 |
| AOC4P + BANCR + CCAT2 + LINC00857 + TINCR | Gastric cancer | Plasma | 0.82 | 0.87 | AUC 0.91 | GAPDH | Zhang et al., 2017 |

N.A. not available / data presented in graphical format in original report

**Table 1. List of blood-based lncRNAs investigated as potential biomarkers for diagnosis of various cancers.** Information reported includes the name of lncRNA, cancer type, source of lncRNA, lncRNA specificity, lncRNA sensitivity, AUC (ROC) value (area under the ROC curve - receiver operating characteristic), QUADAS score, normalization method used in RT-qPCR analyses and literature reference.

| Step | Recommended | Reason | Reference |
|---|---|---|---|
| Patient selection | Exclude patients with inflammation | Higher / different levels of white blood cells associated with inflammation may impact levels of circulating RNAs upon cytolysis | Kroh et al., 2010; Chen et al., 2013 |
| | Recruit patients with same gender, age and race | Minimize variation in lncRNA levels due to possible inter-individual confounding factors (such as SNPs, CNV, etc.) | Tong et al., 2015 |
| | May include questionnaire about diet and lifestyle | Diet and lifestyle (alcohol consumption, smoking) can affect lncRNA levels | Dobosy et al., 2008; Solanas et al., 2009; Thai et al., 2013 |
| Blood sample preparation | Prepare serum or plasma. Discard cellular fraction | Cellular fraction of blood may contain different levels of blood cells which in return may impact levels of circulating RNAs upon cytolysis | Chen et al., 2013; Pritchard et al., 2012 |
| | Strict standard operating procedures when preparing serum / plasma | Minimize variations in circulating RNAs due to sample preparation. Avoid hemolysis. | Pritchard et al., 2012 |
| | Measure $A_{414}$, $A_{541}$, $A_{576}$ | Assess for hemolyzed samples | Permuth et al., 2017 |
| RNA extraction | Use kits compatible with liquid samples | Enable extraction of circulating lncRNAs from plasma or serum samples | Kit manufacturers |
| | Use kits combining both solid (filter) and liquid phase (organic) extraction | Maximize extraction of circulating lncRNAs from plasma or serum samples | Ren et al., 2013; Arita et al., 2013; Svoboda et al., 2014 |
| | Use as much plasma / serum as possible | Maximize RNA yield after extraction | Our recommendation |
| Reverse Transcription | Use same volume of RNA extracts | Allow maximum RNA input for Reverse Transcription | Our recommendation |
| qPCR (relative quantification with ΔΔCt method) | Test several reference genes. Carefully choose best reference gene(s) using NormFinder, RefFinder or Genorm algorithms. Most popular: GAPDH, beta-actin, 18S. To avoid: RPLPO, GUSB, HPRT | The right reference gene is needed for accurate relative quantification using ΔΔCt method. GAPDH, beta-actin, 18S present in large quantities in blood. RPLPO levels inconsistent in blood. GUSB, HPRT levels too low in blood | Gresner et al., 2009; Svoboda et al., 2014; Weber et al., 2013; Dong et al., 2015; Tang et al., 2015; Jing et al., 2016 |
| | Careful in interpretation of data when using spike-in controls | Spike-in controls do not account for variations in lncRNA concentrations in blood-derived samples prior to RNA extraction step | Qi et al., 2016 |
| | Measure transcript levels of MB, NGB, CYGB genes | Assess for contamination from red blood cells | Permuth et al., 2017 |
| | Measure transcript levels of APOE, CD68, CD2, CD3 genes | Assess for contamination from white blood cells | Permuth et al., 2017 |

**Table 2. Guidelines recommended for qPCR-based study of circulating lncRNAs as biomarkers for cancer diagnosis.** Based on troubleshooting performed by previous studies, the information reported here includes step of the analysis, actual recommendation, reason for the recommendation and related literature reference.

**Figure legends**

**Table 1. List of blood-based lncRNAs investigated as potential biomarkers for diagnosis of various cancers**. Information reported includes the name of lncRNA, cancer type, source of lncRNA, lncRNA specificity, lncRNA sensitivity, AUC (ROC) value (area under the ROC curve - receiver operating characteristic), QUADAS score, normalization method and literature reference.

**Table 2. Guidelines recommended for the study of circulating lncRNAs as biomarkers for cancer diagnosis, based on troubleshooting performed by previous works**.
Information reported includes step of the analysis, actual recommendation, reason for the recommendation and related literature reference.

**Figure 1**. **Diagram summarizing the full panel of possible clinical applications that can be derived from the analysis of blood-based lncRNAs**.
Information indicated includes four main domains of applications (cancer prevention, cancer diagnosis, cancer prognosis, cancer treatment) and smaller subdomains referring to the domain of the same color.

**Figure 2. Cancer-specific and multicancerblood-derived lncRNA biomarkers.**
A. Diagram showing circulating lncRNAs reported in the literature regrouped by cancer type. Some lncRNAs (in black letters) are cancer-specific. Other circulating lncRNAs (in white letters) such as MALAT1, SPRY4-IT1, PVT1, UCA1 and LINC00152 reflect tumorigenesis in multiple organs. B. Simplified cartoon representing the specificity of certain circulating lncRNAs towards cancers of organs located in particular anatomic segments of the human body. C. Gene tissue expression of some of the most widely reported circulating lncRNAs with high multi-cancer diagnosis potential (GTEx, obtained from UCSC genome browser).

**Figure 3. Circulating lincRNAs and a common set of protein partners**. A. Data extracted from starBase V2.0 and lncRNome databases reporting lncRNA-protein interactions occurring in tissues. Indicated lncRNAs share the same set of interacting proteins that are also known to be involved in tumorigenesis. These main proteins may constitute an oncogenic pan-lncRNA core protein interactome. Displayed protein-protein interactions are based on data from BioGRID database. B. Graph bars representing the number of interactions with lncRNAs and proteins for each RNA-binding protein shown in (A). C. Putative pan-cancer multimeric RNA-binding protein complex showing the different interactions between the proteins that are the most commonly recruited by cancer-related lncRNAs as shown in (A).